\documentclass[aps,pre,preprint,superscriptaddress]{revtex4-1}

\usepackage{graphicx, amssymb,verbatim, amsmath, siunitx, color,float, soul,makecell,xcolor}
\usepackage{natbib,url, enumerate}
\usepackage[colorlinks, allcolors = blue, bookmarksopen,bookmarksnumbered]{hyperref}

\definecolor{colsq}{rgb}{0,0.4470,0.7410}
\definecolor{coltr}{rgb}{0.8500,0.3250,0.0980}
\definecolor{coldm}{rgb}{0,0.4470,0.7410}
\definecolor{colbtr}{rgb}{0.9290,0.6940,0.1250}

\begin{document}

\title{Symbiotic Dynamics in Living Liquid Crystals}  
\author{Aditya Vats}
\affiliation{Department of Physics, Indian Institute of Technology Delhi, New Delhi -- 110016, India.}
\author{Pradeep Kumar Yadav}
\author{Varsha Banerjee}
\affiliation{Department of Physics, Indian Institute of Technology Delhi, New Delhi -- 110016, India.}
\affiliation{School of Physical Sciences, Jawaharlal Nehru University, New Delhi -- 110067, India.}
\author{Sanjay Puri}
\affiliation{School of Physical Sciences, Jawaharlal Nehru University, New Delhi -- 110067, India.}

\begin{abstract}
An amalgamate of nematic liquid crystals and active matter, referred to as {\it living liquid crystals}, is a promising self-healing material with futuristic applications for targeted delivery of information and micro-cargo. We provide a phenomenological model to study the symbiotic pattern dynamics in this contemporary system using the Toner-Tu model for active matter (AM), the Landau-de Gennes free energy for liquid crystals (LCs), and an experimentally motivated coupling term that favours co-alignment of the active and nematic components. Our extensive theoretical studies unfold two novel steady states, {\it chimeras} and {\it solitons}, with sharp regions of distinct orientational order that sweep through the coupled system in synchrony. The induced dynamics in the passive nematic is unprecedented. We show that the symbiotic dynamics of the AM and LC components can be exploited to induce and manipulate order in an otherwise disordered system.
\end{abstract}

\maketitle

\section{Introduction}

An assembly of interacting particles, ranging from microscopic to macroscopic sizes, that converts energy from the environment into mechanical energy for self-propulsion is termed as active matter (AM). This term encompasses a wide variety of living and non-living systems such as bird flocks, insect swarms, animal herds and fish shoals, suspensions of bacteria, cytoskeletal filaments and protein motors, synthetic self-phoretic colloids, vibrated granular matter, and even human crowds \cite{Ben_1995,Julia_1997,Ndlec_1997,Helbing_2000, Helbing_2000_PRL,Surrey_2001,Hubbard_2004,Sokolov_2007,Shaller_2010,Ramaswamy_2010,Sumino_2012,Wensink_2012,Palacci_2013,Marchetti_2013}. The immense diversity in the constituent particles, lack of time-reversal symmetry and the intrinsic out-of-equilibrium behaviour has lead to intriguing experimental and theoretical investigations, see \cite{Ramaswamy_2010,Marchetti_2013,Bechinger_2016} for different perspectives. Most of these works have discussed AM in isotropic Newtonian fluids. However, recent attention has also turned to AM in non-Newtonian fluids. The fluid endows the active system with unique properties including improved diffusivity and decreased viscosity. An anisotropic medium also introduces directional dependence and can help control the AM's chaotic motion. In this context, a system of great topical interest is that of {\it living liquid crystals} (LLCs), where living (active) particles are introduced in nematic liquid crystals (NLCs) \cite{Zhou_2014,Trivedi_2015,Peng_2016,Lintuvuori_2017,Genkin_2017,Sokolov_2019,Zhou_2018,Turiv_2020}. The latter are classic examples of anisotropic fluids having long-range order (LRO) or quasi-LRO below a critical temperature $T_c$, with a special direction of averaged molecular alignment called the {\it director} ${\bf n}$ \cite{JP_DG_1995,Stephen_1974}. Consequently, mechanical, optical and diffusive properties of NLCs exhibit strong directional dependence \cite{Stephen_1974}. 

The benchmarking works on LLCs considered a low concentration of rod-like bacteria ({\it Bacillus subtilis}) swimmers in non-toxic NLCs confined to a quasi-two-dimensional geometry, and reported spectacular experimental phenomena that were never observed in Newtonian fluids \cite{Zhou_2014,Peng_2016,Genkin_2017,Chi_2020}. The swimming bacteria (flagella) serve as probes for extracting information about the NLC properties and their geometric confinement. They create perturbations in the nematic medium over nanometer scales and yield emergent textures over hundreds of micrometers. The topological defects in NLCs on the other hand, play a critical role in active transport. Experiments reveal that in the defect free regions, the bacteria always swim parallel to the local director.  They accumulate at the T-shape defects (with topological charge +1/2), but are deflected from Y-shape defects (with topological charge -1/2) \cite{Genkin_2017}. Such observations are presumably generic to other self-propelled particles including synthetic swimmers, provided the low-concentration limit is respected. It is believed that LLCs will bridge the properties of active and passive matter to create new micro-fluidic devices that can transport fluids without pumps or pressure, synthetic systems which resemble cells in motion, and nanotechnologies for targeted drug deliveries, sensing and other biomedical applications.

An important direction in this emerging field is to develop models of LLCs so that joint experimental and theoretical efforts can be made to unravel potential applications. One of the first contributions in this direction has been due to Genkin et al. \cite{Genkin_2017}, who introduced continuum models that capture the experimentally observed pattern formation of rod-shaped bacteria in NLCs. Guided by experimental observations, the primary assumptions in the description of Genkin et al. are: (i) The volume fraction of bacteria is relatively low and does not perturb the properties of the suspending NLC; (ii) The suspended bacteria co-align with the local nematic director on a time-scale much smaller than the characteristic time of collective behavior; (iii) At each point in the quasi-two-dimensional space, interactions between bacteria are apolar and allow them to glide past without collisions. To model the NLC environment, Genkin et al. use the Berris-Edwards model comprising of equations of motion for the tensor order parameter field ${\bf Q}({\bf r},t)$ and the velocity field ${\bf u}({\bf r},t)$. The transport of bacteria is governed by two coupled advection-diffusion equations for the concentrations of bacteria swimming parallel $c^+$ and anti-parallel $c^-$  to the director ${\bf n}$ \cite{Genkin_2017,Harvey_2013,Shi_2014}. This model reproduces the experimentally observed accumulation and expulsion of bacteria at the defect cores. The above work is of great interest but is restricted to the dilute regime, where bacteria do not directly interact with each other. Clearly, the dense limit is significant in many applications of AM. Moreover, the pioneering experiments of Zhou et al. \cite{Zhou_2014} on LLCs showed a rich and fascinating phenomenology in this limit also.

The scope of AM is vast. It studies the collective behaviour of self-propelled particles of varying sizes in a plethora of environments. The interaction of active particles amongst themselves, and with the medium, can be expected to yield exotic dynamical patterns with novel applications. An important direction of research therefore is to construct generic models of LLCs that capture pattern formation for the case when all three interactions are significant: AM-AM, LC-LC and AM-LC. In this situation, we expect a symbiotic dynamics with complex interplay of AM and LCs. We embark on this path by considering two well-established coarse-grained descriptions, the Toner-Tu (TT) model for AM and the Landau-de Gennes (LdG) free energy for NLCs, along with a coupling term motivated by experimental observations \cite{Genkin_2017}. The LdG formulation does not incorporate hydrodynamics, so there is no inherent director dynamics. The latter is usually imparted by the coarse-grained time-dependent Ginzburg-Landau (TDGL) equations, and is purely relaxational \cite{Puri_2009,Bray_2002}.

Our extensive simulations reveal two novel steady states in the LLCs: (i) Sharp bands of large orientational order (in AM and NLCs) coexisting with a background of disoriented AM and isotropic NLCs. We refer to this coexistence of order and disorder as a {\it chimera} state, a term which has found usage in the nonlinear dynamics literature \cite{Kuramoto_2002,Abrams_2004}. The bands sweep through the system with the speed of the active particles (say $v_0$). The band-width $\Delta$ exhibits a power-law dependence on the AM-NLC coupling: $\Delta \sim (c_0^* - c_0)^{\theta}$, where $\theta$ is a universal exponent. (ii) Localized regions with large orientational order (in AM as well as NLCs) or {\it solitons} that propagate with speed $v_0$. There are several 1-dimensional equations \cite{Newell_1985,Dickey_2003,SPuri_1990} which are known to exhibit soliton solutions, i.e., solitary waves which maintain their integrity under collision with other solitary waves. These are ubiquitous in diverse physical systems, ranging from plasmas to fluids and nerve conduction. However, there are very few examples of solitons in dimensions higher than 1. The simulations of our model for LLCs show four kinds of steady states: chimera, soliton, ordered and disordered. We have evaluated the phase boundaries analytically from the fixed points of the dynamical equations and their linear stability analysis. 

\section{Model and Theoretical Framework}

Deep insights on NLCs have emerged from mean-field approaches based on the minimization of the LdG free energy  \cite{JP_DG_1995,Mottram_2014}. This is obtained as a Landau expansion in terms of a mesoscopic order parameter ${\bf Q}$, and is characterized by a few phenomenological constants. The ${\bf Q}$-tensor is symmetric and traceless, with elements $Q_{ij}=\mathcal{S}\left(n_in_j - \delta_{ij}/2\right)$. The eigenvector corresponding to the largest eigenvalue is the director ${\bf n}$, and $\mathcal{S}$ measures the orientational order about ${\bf n}$. The isotropic phase ($T>T_c$) corresponds to $\mathcal{S}=0$, and $\mathcal{S}=1$ describes the fully aligned nematic phase ($T<T_c$). A defect corresponds to regions of low order or $\mathcal{S} \simeq 0$. It is easy to check that, in $d=2$
\begin{equation}
\text{Tr}({\bf Q}) =0; \quad \text{Tr}({\bf Q}^2) =2(Q_{11}^2+Q_{12}^2) = \mathcal{S}^2/2; \quad \text{Tr}({\bf Q}^3) =0 .
\end{equation}

The LdG free energy for NLCs has been modelled as \cite{JP_DG_1995,Mottram_2014}
\begin{eqnarray}
\label{LdG_FE}
F_Q[{\bf Q}] &=& \int \mbox{d}{\bf r} \left\{ \frac{A}{2}\mbox{Tr}(\boldsymbol{Q}^2)+\frac{B}{3}\mbox{Tr}(\boldsymbol{Q}^3)+\frac{C}{4}[\mbox{Tr}(\boldsymbol{Q}^2)]^2 + \frac{L}{2}\left|\nabla{\bf Q}\right|^2\right\}.
\end{eqnarray}
The Landau coefficients $A,B,C$ and $L$ are phenomenological parameters which are related to experimentally determined quantities like critical temperature, latent heat of transition, magnitude of the order parameter, etc. \cite{Priestly_2012,Hberg_2015}. For example, $A=A_0(T-T_c)$, where $A_0$ is a material dependent coefficient and $T_c$ is the critical temperature. At the coarse-grained level, the appropriate framework to study the dissipative dynamics that drives the system to the free energy minimum is the TDGL equation \cite{Puri_2009,Bray_2002}:
\begin{equation}
\frac{\partial {{\bf Q}}}{\partial t} = -\Gamma_{{\bf Q}} \frac{\delta F_Q[\mathbf{Q}]}{\delta {{\bf Q}}}.
\label{tdgl}
\end{equation}
The parameter $\Gamma_Q$ is the damping factor for the nematic component and sets the relaxation time scale for the system. The terms on the right of Eq.~\eqref{tdgl} are the functional derivatives of the free energy functional. 

The minimal microscopic description for the collective motion of AM is the Vicsek model \cite{Vicsek_1995}. The corresponding coarse-grained formulation, provided by the elegant hydrodynamic theory of Toner and Tu (TT), yields the equation of motion for (i) the local density of the active particles $\rho(\bf{r},t)$, and (ii) the local polarization ${\bf P}({\bf r},t)$ that describes their average orientation \cite{Toner_1995,Toner_1998,Ramaswamy_2010,Mishra_2010,Marchetti_2013}. 
Although the original model is formulated phenomenologically using symmetry considerations, it is instructive to rewrite the equations of motion in terms of a free energy functional $F_a[\rho, {\bf P}]$ \cite{Ramaswamy_2010,Marchetti_2013}:
\begin{eqnarray}
 \frac{\partial \rho}{\partial t}&=& -v_0 \nabla \cdot({\bf P}\rho) - \nabla \cdot \left(-\Gamma_{\rho}\nabla \frac{\delta F_a}{\delta \rho}\right) , \label{DD} \\
\frac{\partial {\bf P}}{\partial t}&=& \lambda_1 ({\bf P} \cdot \nabla ){\bf P} - \Gamma_{P}\frac{\delta F_a}{\delta {\bf P}} . \label{DP}
\end{eqnarray}
Here, $v_0$ is the speed of the active particles, and $\Gamma_{\rho}$ and $\Gamma_P$ set the relaxation time scales for the density and polarization fields. The first term in Eq.~\eqref{DD} quantifies the change in the density due to the polarization field. In the TT model, the ${\bf P}$-field acts both as the current and the orientational order parameter. Hence, it evolves in time [Eq.~(\ref{DP})] via both advection and flow alignment. Further, $\lambda_1$ has the dimension of the speed and Galilean invariance would require $\lambda_1=v_0$. Since this is a non-equilibrium system, $\lambda_1$ is generally a phenomenological parameter different from $v_0$. 

The free energy functional in Eqs.~\eqref{DD}-\eqref{DP} is given by \cite{Ramaswamy_2010,Marchetti_2013}:
\begin{eqnarray}
F_a[\rho, {\bf P}] = \int \text{d}{\bf r} \left[\frac{\alpha(\rho)}{2}|{\bf P}|^2+\frac{\beta}{4}|{\bf P}|^4 +\frac{\kappa}{2}|\nabla {\bf P}|^2+ \frac{w}{2}|{\bf P}|^2\nabla\cdot {\bf P} -\frac{v_1}{2} (\nabla\cdot{\bf P}) \frac{\delta \rho}{\rho_0}+ \frac{D_\rho}{2}({\delta \rho})^2\right] , \label{fa}
\end{eqnarray}
where $\alpha, \beta, \kappa,  w, v_1, D_{\rho}$ are material-dependent parameters whose precise values can be related to the microscopic properties of the active particles \cite{Bertin_2006,Bertin_2009}. The parameter $\alpha(\rho)=\alpha_0 (1-\rho/\rho_c)$, where $\rho_c$ is the critical density that is required to observe order in the active system. The gradient term $|\nabla {\bf P}|^2$ models the energy cost for a deformation of the order parameter. The next two terms in the equation provide the $|{\bf P}|^2$ and density contributions to the spontaneous splay $\nabla \cdot {\bf P}$. These terms can be interpreted as the local aligning field due to the density and orientational order $|{\bf P}|^2$. The last term in Eq.~(\ref{fa})  penalizes the variation in the density about its mean value: $\delta \rho = \rho - \rho_0$. A detailed discussion of these terms and their applicability can be found in \cite{Ramaswamy_2010} and \cite{Marchetti_2013}. 

Some remarks about the states seen in the TT model are in order. The order-disorder transition takes place as the parameter $\alpha(\rho)$ goes through zero. An average density $\rho_0<\rho_c$ results in a {\it disordered phase} with ${\bf P}=0$. For $\rho_0>\rho_c$, the system shows a state of uniform orientational order with $|{\bf P}|^2 \sim (\rho_0 /\rho_c-1)$. This {\it ordered phase} is characterized by the movement of active particles with velocity ${\bf v}=v_0{\bf P}$. Near the transition point ($\rho_0=\rho_c^+$), the ordered phase is unstable, and the system relaxes to a {\it banded phase} that sweeps through the system with speed $v_0$ \cite{Ramaswamy_2010,Mishra_2010}. Additionally, solitons have also been observed in the quasi-one-dimensional case, but not in higher dimensions  \cite{Bertin_2009,Ihle_2013,Toner_2014}. 

The above coarse-grained models are the ingredients of our phenomenological model for LLCs. We write the free energy of this composite system as the sum of (a) free energies of the nematic and active components, and (b) a suitably designed coupling term. Keeping in mind the experimental observations of Genkin et al. \cite{Genkin_2017}, we define the coupling between the nematic and active component as the dyadic product of the ${\bf Q}$-tensor and the polarization vector ${\bf P}$. This is the lowest order term that ensures ${\bf P} \parallel {\bf n}$ \cite{konark_2019,Konark_2019_2,Aditya_2020,Aditya_2021,Aditya_2022}. With these considerations, the free energy for the LLC can be written as
\begin{equation}
F[{\bf Q}, \rho, {\bf P}] =F_a+F_Q - c_0 \sum_{i,j} { Q_{ij}}P_i P_j,
\label{fe}
\end{equation}
where $c_0$ quantifies the strength of the AM-nematic interaction. Note that, when stated in terms of ${\bf n}$, the coupling term takes the form $-({\bf n} \cdot {\bf P})^2$, which makes it easy to see that the two components prefer co-alignment \cite{konark_2019,Konark_2019_2,Aditya_2020,Aditya_2021,Aditya_2022}. 

We now substitute the free energy defined in Eq.~\eqref{fe} in Eqs.~\eqref{tdgl}-\eqref{DP}, and retain gradient terms up to second order to obtain the dynamical equations for LLCs in $d=2$. These are provided in Eqs.~\eqref{TDGL_lc1}-\eqref{eq_rho} of Appendix \ref{Appendix_1}. Note that our model, which does not include the hydrodynamics of the nematic matrix, is suitable when the AM-nematic interactions are short-ranged, and the velocity of the nematogen is small as compared to the propulsion velocity of the active particle. This is the case in Ref.~\cite{Toner_2014}, or for AM in pre-designed director patterns \cite{Peng_2016,Turiv_2020}.

The dimensionless form of Eqs.~\eqref{TDGL_lc1}-\eqref{eq_rho} can be obtained by introducing the rescaled variables 
\begin{equation}
{\bf Q} = c_Q{\bf Q}^\prime, \quad {\bf P} = c_P{\bf P}^\prime, \quad {\bf r} =c_r {\bf r}^\prime, \quad t= c_t t^\prime .
\end{equation}
The appropriate scale factors are
\begin{equation}
c_Q = \sqrt{\frac{|A|}{2C}}; \quad c_P=\sqrt{\frac{\alpha_0}{\beta}}; \quad c_t=\frac{\beta}{\alpha_0\Gamma_Q} \sqrt{\frac{|A|}{2C}}; \quad
c_r=\sqrt{\frac{L}{|A|}} .
\end{equation}
Dropping the primes on the variables, we obtain
\begin{eqnarray}
\frac{\partial Q_{11}}{\partial t}&=&\xi_1\left[\pm Q_{11}-(Q_{11}^2+Q_{12}^2)Q_{11}+\nabla^2Q_{11}\right]+c_0 (P_1^2-P_2^2), \label{eq_Q1} \\
\frac{\partial Q_{12}}{\partial t}&=&\xi_1 \left[\pm Q_{12}-(Q_{11}^2+Q_{12}^2)Q_{12}+\nabla^2Q_{12}\right]+ 2c_0 P_1P_2,\label{eq_Q2} \\
\frac{1}{\Gamma}\frac{\partial P_1}{\partial t}&=& \xi_2 \bigg[\left(\frac{\rho}{\rho_c}-1-{\bf P}\cdot {\bf P}\right) P_1 - \frac{v_1^\prime}{2\rho_0} \nabla_x \rho + \lambda_1^\prime ({\bf P} \cdot \nabla)P_1 +\lambda_2^\prime \nabla_x (|{\bf P}|^2) \nonumber \\
&& + \lambda_3^\prime P_1(\nabla \cdot {\bf P}) + \kappa^\prime \nabla^2 P_1 \bigg] + c_0 (Q_{11}P_1+Q_{12}P_2), \label{eq_P11} \\
\frac{1}{\Gamma}\frac{\partial P_2}{\partial t}&=&\xi_2 \bigg[\left(\frac{\rho}{\rho_c}-1-{\bf P}\cdot {\bf P}\right) P_2 - \frac{v_1^\prime}{2\rho_0} \nabla_y \rho + \lambda_1^\prime({\bf P} \cdot \nabla)P_2 +\lambda_2^\prime \nabla_y (|{\bf P}|^2) \nonumber \\
&& +\lambda_3^\prime  P_2(\nabla \cdot {\bf P}) + \kappa^\prime\nabla^2 P_2\bigg] + c_0 ( Q_{12}P_1-Q_{11}P_2), \label{eq_P22} \\
\frac{1}{\Gamma^\prime}\frac{\partial \rho}{\partial t}&=& -v_0^\prime \nabla\cdot({\bf P}\rho) + D_{\rho}^\prime \nabla^2 \rho. \label{Rho}
\end{eqnarray}
The dimensionless parameters in Eqs.~\eqref{eq_Q1}-\eqref{Rho} are:
\begin{eqnarray}
&& \xi_1 = \dfrac{2|A|\beta}{\alpha_0}\sqrt{\dfrac{|A|}{2C}}, \quad
\xi_2=\dfrac{\alpha_0}{2} \sqrt{\dfrac{2C}{|A|}} , \nonumber \\ 
&& v_1^\prime=\dfrac{v_1}{\alpha_0}\sqrt{\dfrac{\beta|A|}{\alpha_0L}}, \quad v_0^\prime = \dfrac{v_0}{\Gamma_{\rho}}\sqrt{\dfrac{\alpha_0|A|}{\beta L}}, \nonumber \\
&& \Gamma=\dfrac{\beta|A|\Gamma_P}{\alpha_0\Gamma_QC}, \quad
\Gamma^\prime=\dfrac{\beta\Gamma_{\rho}}{\alpha_0\Gamma_Q}\sqrt{\dfrac{|A|}{2C}}, \nonumber \\
&& \kappa^\prime=\dfrac{\kappa |A|}{\alpha_0 L}, \quad
D_{\rho}^\prime=\dfrac{D_{\rho} |A|}{L}, \nonumber \\
&& \lambda_1^\prime=\dfrac{\lambda_1}{\Gamma_P}\sqrt{\dfrac{|A|}{\alpha_0\beta L}},  \quad \lambda_2^\prime= \lambda_2 \sqrt{\dfrac{|A|}{\alpha_0\beta L}}, \quad 
\lambda_3^\prime= \lambda_3 \sqrt{\dfrac{|A|}{\alpha_0\beta L}} .
\end{eqnarray}  

The $\pm$ sign in Eqs.~\eqref{eq_Q1}-\eqref{eq_Q2} determines whether the nematic component (in the absence of AM) is above ($-$) or below ($+$) its critical temperature $T_c$. Before presenting results, let us discuss the choice of parameters. The quantities $\xi_1$ and $\xi_2$ depend on the relative magnitudes of ${\bf Q}$ and ${\bf P}$, and are set to 1 in our simulations. In dimensional units, $v_0 >0$ is the speed of the active particle. Further, the stable state exists only if $v_1 >0$ \cite{Bertin_2009}. We assign the corresponding recaled parameters the values $v_0^\prime = 0.5, v_1^\prime = 0.25$. Our simulation results do not change significantly if $v_0^\prime, v_1^\prime$ are varied. The dimensional parameters $\Gamma_P, \Gamma_Q$ and $\Gamma_\rho$ are the inverse relaxation scales of ${\bf P}, {\bf Q}$ and $\rho$, respectively. The dimensionless quantities $\Gamma$ and $\Gamma^\prime$ measure the relative time-scales, and we set them to 1. Similarly, $\kappa^\prime$ and $D_\rho^\prime$ set the relative values of elastic scales, and we assign them the value 1. Finally, the $\lambda_i$ are the strengths of the convective nonlinearities present due to the absence of Galilean invariance. As remarked in Appendix \ref{Appendix_1}, the terms with $\lambda_2$ and $\lambda_3$ arise from the same term in the free energy $F_a$ and obey $\lambda_2 = -\lambda_3/2$. However, both these terms are allowed under symmetry considerations, and we treat $\lambda_2$ and $\lambda_3$ as independent parameters. In dimensional terms, the linear stability analysis of the TT equations shows that non-trivial states arise under the conditions $\lambda_1/\Gamma_P + \lambda_2 + \lambda_3 < 0$ and $\lambda_2=-\lambda_3$ \cite{Ramaswamy_2010,Marchetti_2013}. These conditions are invariant under the above rescaling, and we consider the case with $\lambda_1^\prime =-0.5, \lambda_2^\prime =-0.5, \lambda_3^\prime =0.5$. There is clearly a degree of freedom involved in the above choice of parameters. However, we emphasize that our numerical results do not change qualitatively on changing the above values as long as the specified signs are preserved. The coupling constant $c_0$ will be allowed to vary in our simulations.

\section{Results}

At the core of the current theoretical modelling is to understand the interplay of the AM-NLC coupling in LLCs. We now focus on understanding the effect of the coupling strength $c_0$ on the dynamical evolution of the active and nematic fields. The three cases which provide interesting outcomes are {\it Case 1}:  $T>T_c$, $\rho_0 = \rho_c^+$; {\it Case 2}: $T<T_c$, $\rho_0 = \rho_c^-$; {\it Case 3}: $T<T_c$, $\rho_0 = \rho_c^+$. Here, $\rho_c^+$ ($ \rho_c^-$) corresponds to density slightly above (below) the critical density $\rho_c$.  Without loss of generality, we choose $\rho_c=0.5$. For each of the three cases, we numerically solve Eqs.~\eqref{eq_Q1}-\eqref{Rho} via Euler discretization with an isotropic Laplacian on an $N^2$ lattice ($N = 128$). We impose periodic boundary conditions in both directions \cite{kincaid_2009}, so as to remove the edge effects and mimic the bulk system. The discretization mesh sizes are chosen to be $\Delta t = 0.01$ and $\Delta x = 1.0$. The initial conditions for ${\bf Q}$ and ${\bf P}$ are chosen as small fluctuations about zero, which mimics the disordered state. The corresponding initial state for $\rho$ is small fluctuations around the mean density $\rho_0$. All statistical quantities have been averaged over 10 independent initial conditions, unless otherwise stated.

\begin{figure}[H]
\centering
\includegraphics[width=0.9\linewidth]{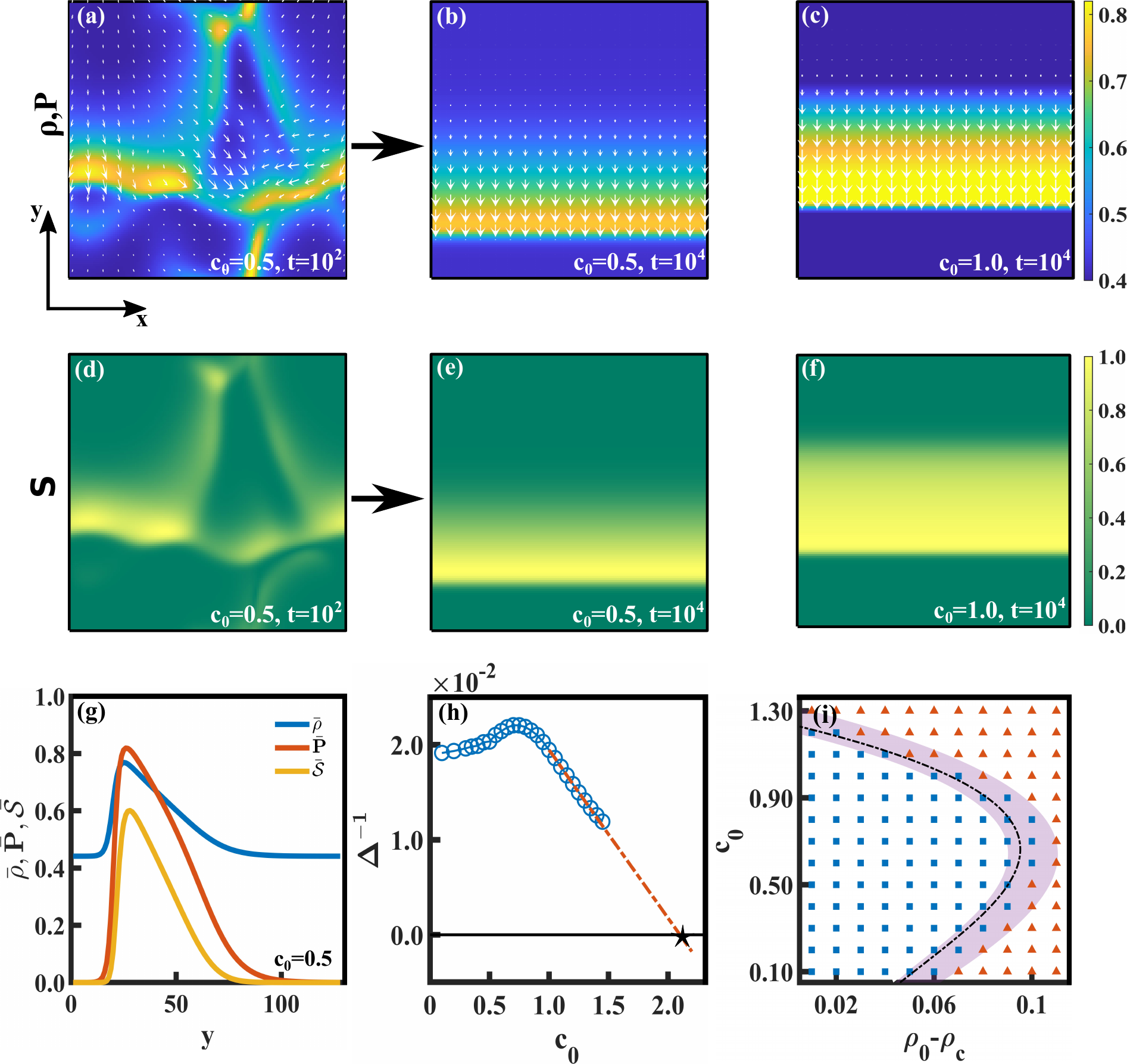} 
\caption{Morphology snapshots for the active field (first row) and nematic field (second row) in {\it Case 1} ($T > T_c, \rho_0=\rho_c^+ =0.52$) for specified values of ($t,c_0$). The color bar in the top row indicates the density ($\rho$) of the active field; the white arrows represent the direction and magnitude of the polarization field (${\bf P}$). The colour bar in the second row shows the orientational order $\mathcal{S}$ in the nematic, see text for details. Sub-figure (g) shows the variation of $\bar{\rho}$, $\bar{{\bf P}}$ and $\bar{{\mathcal S}}$ with $y$ for morphologies (b) and (e), where the bar indicates an average along the $x$-direction. Sub-figure (h) shows the dependence of the inverse band width $\Delta^{-1}$ on the coupling $c_0$. The dashed line corresponds to $\Delta^{-1} = c_0^*-c_0$, with $c_0^* = 2.1$. Sub-figure (i) shows the phase diagram demarcating the ordered ({\color{coltr}$\blacktriangle$}) and chimera ({\color{colsq}$\blacksquare$}) states. The dashed line indicates the analytical phase boundary obtained in Appendix \ref{Appendix_2}, while the smeared region indicates the approximate numerical counterpart. The smeared region will reduce to the analytical results for infinite system size and $\Delta x, \Delta t \rightarrow 0$.}
\label{f1}
\end{figure}

First, let us discuss the consequences of AM-LC coupling for {\it Case 1}. The linear stability analysis for the uncoupled system ($c_0=0$) yields a disordered state for the nematic component with $\mathcal{S}\simeq 0$, and a banded state for the active component. Fig.~\ref{f1} shows the evolution of the active and nematic components with $\rho_0=\rho_c^+=0.52$ for different values of $c_0$. Sub-figures (a) and (b) show the density (see colour bar) of the active field at $t=10^2$ and $10^4$ for $c_0=0.5$. The white arrows point along the ${\bf P}$-field with the length proportional to the magnitude. Clearly, the AM shows a banded state analogous to the uncoupled limit. In the banded state, there is coexistence of order (large $P$) and disorder (small $P$) in the ${\bf P}$-field. In the nonlinear dynamics literature, this has often been referred to as a {\it chimera} state \cite{Kuramoto_2002,Abrams_2004}. In Figs.~\ref{f1}(a)-(b), the evolution to the chimera state is evident. The chimera sweeps through the system with velocity $v_0$. The corresponding developments in the nematic field are shown in Fig.~\ref{f1}(d)-(e). The colour bar indicates the value of the orientational order parameter $\mathcal{S}$, which has been normalized by its maximum value: $\mathcal{S}_{m} \simeq 0.67$ in (d), $\mathcal{S}_{m} \simeq 0.61$ in (e). The coupling imprints the chimera state on the nematic component, which also travels with speed $v_0$. Note that the nematogens continue to remain passive, it is only the orientational order (and disorder) that is dynamical.  A visualization of this novel LLC steady state is provided by Movie 1 of Appendix \ref{Appendix_3}. In Fig.~\ref{f1}(g), we have plotted the variation of $\bar{\rho}$, $\bar{P}$ and $\bar{\mathcal{S}}$ with $y$ in the steady state. The bar indicates an average along the $x$-direction. The homologous variation of all the quantities confirms their spatial co-alignment. These solutions  correspond to traveling waves of Eqs.~\eqref{eq_Q1}-\eqref{Rho} with speed $v_0$. The resultant ordinary differential equations have to be solved numerically to obtain the inhomogeneous profiles in Fig.~\ref{f1}(g).

To examine the consequence of increasing coupling strength, we show the  active and nematic fields for $c_0=1.0$ at $t=10^4$ in sub-figures (c) and (f). The band width ($\Delta$) broadens, and the orientational order increases ($\mathcal{S}_{m} \simeq 1.79$). Sub-figure (h) shows the dependence of $\Delta^{-1}$ vs. $c_0$. The system settles to a homogeneous state ($\Delta^{-1}=0$) at a critical value $c_0^*\simeq 2.1$. The dashed line corresponds to $\Delta^{-1} = c_0^* - c_0$, and is a good fit to the data for higher values of $c_0$. (We attribute the discrepancy in the value of $c_0^*$ to finite system sizes used in our simulations.) In sub-figure (i), we provide the phase diagram in the $(c_0,\rho_0$) plane depicting regions where the chimera and ordered states are stable solutions. We have obtained the phase boundary (dashed line) analytically using linear stability analysis, the details of which are provided in Appendix \ref{Appendix_2}. The smear indicates the region where the numerically obtained phase boundary lies. In this region, the final state obtained in our simulations is dependent on the initial condition and may be either chimera or ordered. This ambiguity is a consequence of the Euler discretization on finite lattices, and will go away for infinite system size and $\Delta x, \Delta t \rightarrow 0$. In the latter limit, we will recover the analytical phase boundary. It should be noted that there is a re-entrant phase transition for a range of $\rho_0$-values, where the LLC makes a transition from ordered $\rightarrow$ chimera $\rightarrow$ ordered on increasing $c_0$. 

\begin{figure}[H]
\centering  
\includegraphics[width=0.9\linewidth]{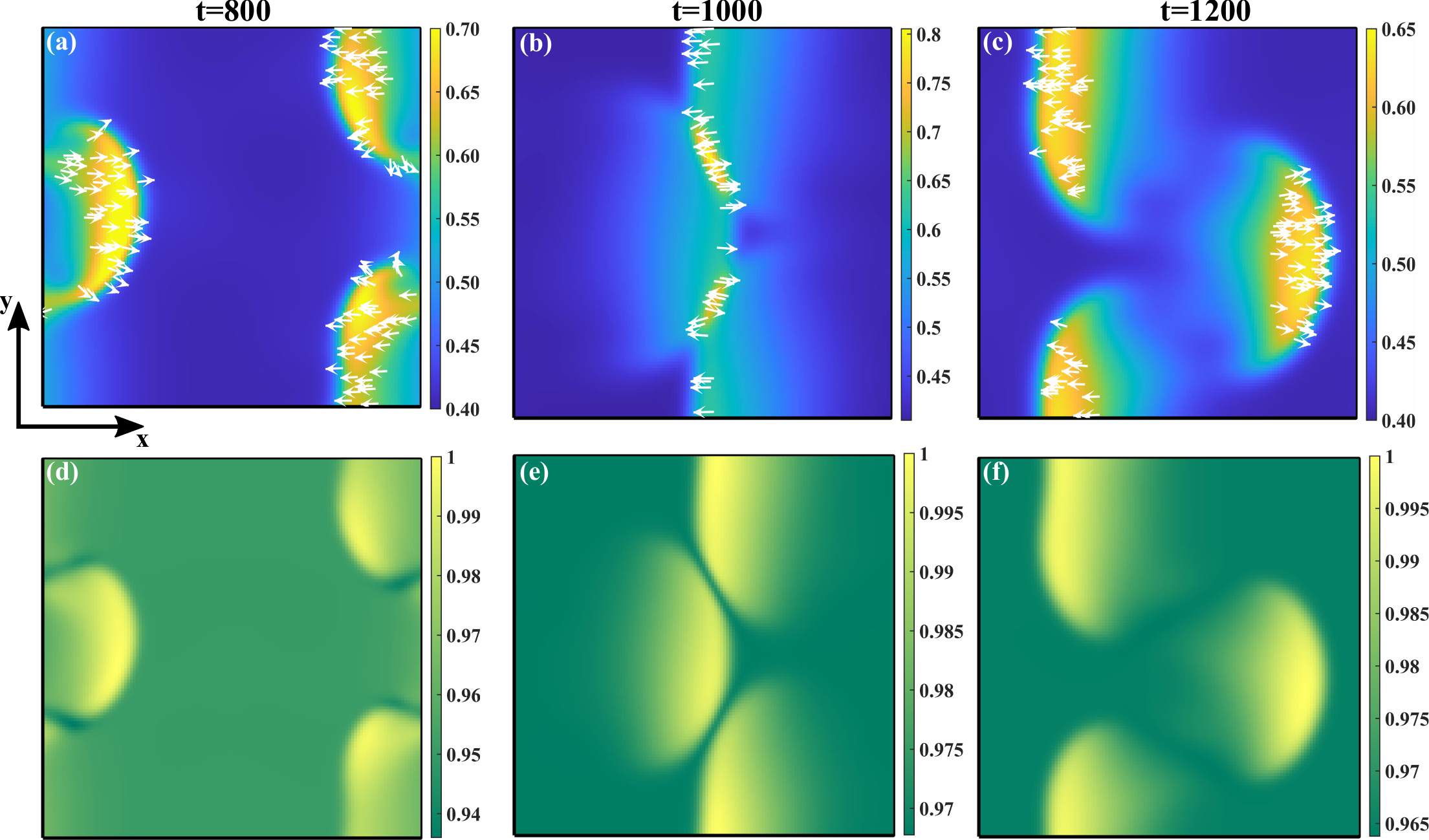}
\caption{Morphology snapshots of the active field (top row) and nematic field (bottom row) for {\it Case 2} with $T<T_c, \rho_0=\rho_c^-=0.48$ and $c_0=0.1$. The arrows in the active morphologies correspond to the polarization field in the high density regions ($\rho>0.6$), and denote the direction of motion of the active field. The $\mathcal{S}$-field is normalized by (d) $\mathcal{S}_m=2.104$, (e) $\mathcal{S}_m=2.066$, (f) $\mathcal{S}_m= 2.0737$ respectively.}
\label{f2}
\end{figure}

Next, we present the results for {\it Case 2} with $T<T_c$, $\rho_0 = \rho_c^- = 0.48$. In the uncoupled limit ($c_0=0$), the ${\bf Q}$-field settles to an ordered nematic state with a non-zero value of $\mathcal{S}$, and the $\rho$ and ${\bf P}$ fields are isotropic. The introduction of the coupling shows dramatic consequences. The active field evolves into a chimera which has so far been observed only when $\rho_0=\rho_c^+$. The naturally ordered nematic state is also driven into a chimera. A prototypical evolution can be seen in Movie 2 of Appendix \ref{Appendix_3}. Additionally, we also observe elusive 2-dimensional {\it soliton} structures for some choices of $c_0$ and $\rho_c^-$. (The probability of occurrence of solitons is around $0.1$ in our simulations.) As mentioned earlier, there is a long history of soliton solutions in completely integrable partial differential equations \cite{Newell_1985,Dickey_2003,SPuri_1990}. Most known soliton equations (e.g., Korteweg-de Vries equation, nonlinear Schrodinger equation, etc.) are 1-dimensional, and there are very few examples of higher-dimensional solitons. We observe these in our proposed model of LLCs. In Fig.~\ref{f2}, we have plotted the evolution of the $\rho$ field (top row) and nematic field (bottom row) for $c_0=0.1$ at $t=800, 1000, 1200$. The white arrows in the active morphologies correspond to the polarization field in the high density regions ($\rho>0.6$). A localized lump ($L_1$) moves to the right ($t=800$), and undergoes a complicated nonlinear collision with lumps moving towards the right ($t=1000$). After this collision, $L_1$ emerges and recovers its original profile. Thus, the solitons maintain their self-confined shapes while propagating and survive the collisions. This scenario can be seen clearly in Movie 3 of Appendix \ref{Appendix_3}. The LLC model proposed here is a dissipative system and not Hamiltonian. So the conventional explanation of soliton behavior via ``complete integrability and infinite constants of motion'' does not apply here. Clearly, the origin of this soliton-like behavior requires further analytical investigation, and is beyond the scope of this paper.

\begin{figure}[H]
\centering  
\includegraphics[width=0.9\linewidth]{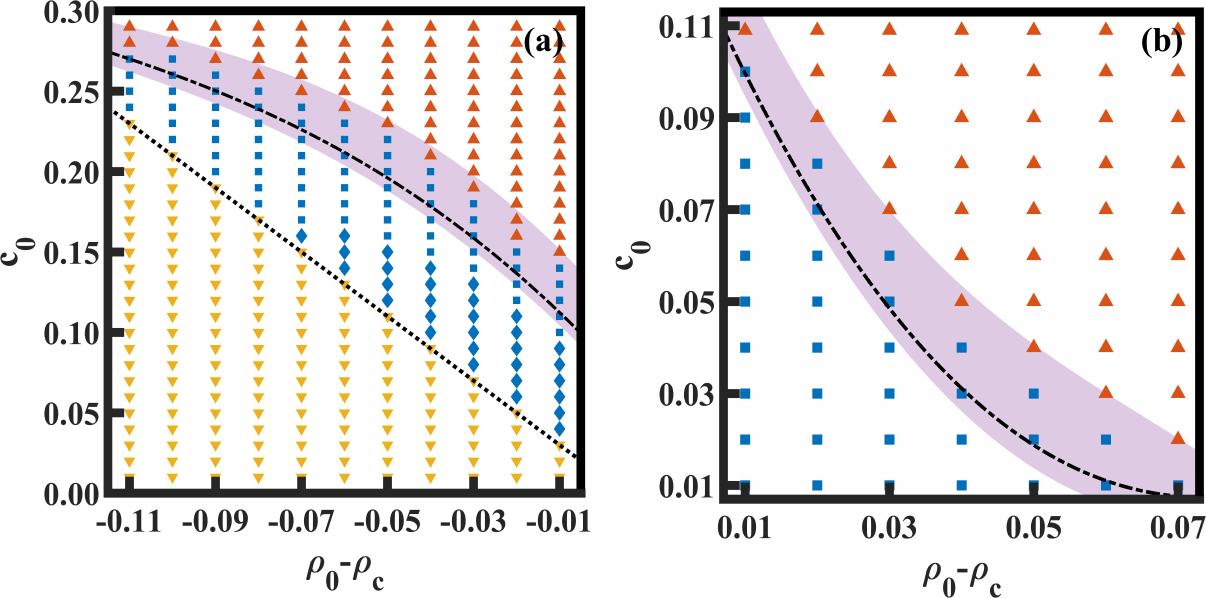}
\caption{Phase diagram for (a) {\it Case 2}: $T<T_c, \rho_0=\rho_c^-$; and (b) {\it Case 3}: $T<T_c, \rho_0=\rho_c^+$ showing different phases: disordered ({\color{colbtr}$\blacktriangledown$}), chimera ({\color{colsq}$\blacksquare$}), soliton plus chimera ({\color{coldm}$\blacklozenge$}), and ordered ({\color{coltr}$\blacktriangle$}). The phase boundaries shown by the dotted and dashed lines are obtained analytically in Appendix \ref{Appendix_2}. The smear indicates the corresponding numerical phase boundary for the chimera $\rightarrow$ ordered transition.}
\label{f3}
\end{figure}

Finally, we present the phase diagrams for {\it Case 2} and {\it Case 3} in Fig.~\ref{f3}(a)-(b) respectively. For {\it Case 2} [Fig.~\ref{f3}(a)], the LLC coupling drives the active system from a disordered state to structured steady states even though $\rho_0=\rho_c^-$. From our linear stability analysis provided in Appendix \ref{Appendix_2}, the transition from the disordered to ordered state occurs when $c_0 + \rho/\rho_c - 1 > 0$, shown by the dotted line. For intermediate values of $c_0$, there is a small region exhibiting both 1-dimensional chimera and higher-dimensional soliton states, and another where only the chimera state is observed. For larger $c_0$-values, the ordering nematic drives AM and both sub-systems transit to an ordered state. For {\it Case 3} [Fig.~\ref{f3}(b)], the nematic and active fields are both in the ordered state with $T<T_c$ and $\rho_0=\rho_c^+$. The region corresponding to chimera states diminishes as ($\rho_0-\rho_c$) increases. For large $c_0 > c_0^*(\rho_0)$, the system transits to an ordered state. In both sub-figures, the dashed line is the analytical phase boundary obtained from the linear stability analysis provided in Appendix \ref{Appendix_2}. The smear, as mentioned earlier, indicates the location of the approximate phase boundaries from our numerics.     

\section{Summary and Conclusion}

To summarize, we have explored pattern dynamics in living liquid crystals (LLCs) - an amalgamate of active matter (AM) and nematic liquid crystals (NLCs). The latter are classic examples of anisotropic materials with a special direction of average molecular alignment. We model the LLCs using the Toner-Tu (TT) model, the Landau-de Gennes (LdG) free energy and an experimentally motivated coupling term that favours co-alignment of the local polarization in the active field and the nematic director. The early theoretical models for this contemporary system are restricted to the dilute regime where the active particles do not interact with one another. Our generic model on the other hand, includes AM-AM, NLC-NLC as well as AM-NLC interactions, which unfold novel symbiotic dynamics of the active and nematic components. 

We focus on understanding this symbiotic dynamics in two-dimensional ($d=2$) LLCs. Such geometries have been realised experimentally in the context of pure NLCs confined to shallow wells by ensuring that the top and bottom surfaces enforce planar boundary conditions. Consequently, the nematic molecules are primarily confined in a plane and the variations along the height of the sample are negligible. Our benchmarking work yields a range of analytical and numerical results for $d=2$ LLCs. From a fixed point analysis of the dynamical equations, we have obtained phase diagrams for a range of parameters. Our extensive theoretical studies unfold two steady states hitherto unobserved in LLCs: (i) {\it Chimeras} corresponding to bands of large orientational order (in AM and NLCs) coexisting with disorder. The ordered regions in the two components are co-aligned, and sweep through the system in synchrony with the speed $v_0$ of the active particles. (ii) {\it Solitons} corresponding to localized regions of order (in AM and NLCs) which are robust under locomotion and collisions. While their presence in $d=1$ is well known, the existence of solitons in higher dimensions is rare. The induced dynamics in the passive nematic is unprecedented. 

Our theoretical framework demonstrates  that the AM-LC coupling can discipline AM by inducing orientational order and heal NLCs by erasing topological defects. Such observations suggest the design and synthesis of new self-healing materials, which can also provide targeted delivery of information and micro-cargo without channels. Our work provides many ideas for manipulating AM and LCs for exciting futuristic applications. We hope that it will initiate joint experimental and theoretical investigations in the contemporary LLCs.

\section{Author contributions}

VB and SP formulated the problem. AV and PY performed the numerical simulations. AV, PY, VB and SP did the analysis and wrote the paper.

\section{Acknowledgements}

AV and PY acknowledge UGC, India for support via a research fellowship. VB acknowledge DST India for research grants. AV and VB gratefully acknowledge the HPC facility of IIT Delhi for computational resources.

\newpage
\appendix
\section{Dynamical Model for LLCs}
\label{Appendix_1}

We substitute the free energy defined in Eq.~(7) in Eqs.~(3)-(5), and keep gradient terms up to second order to obtain the following model for LLCs in $d=2$:
\begin{eqnarray}
\frac{1}{\Gamma_Q}\frac{\partial Q_{11}}{\partial t}&=&\pm 2|A|Q_{11} - 4C(Q_{11}^2+Q_{12}^2)Q_{11} + 2L\nabla^2Q_{11} + c_0 (P_1^2-P_2^2), \label{TDGL_lc1} \\
\frac{1}{\Gamma_Q}\frac{\partial Q_{12}}{\partial t}&=& \pm 2|A|Q_{12}-4C(Q_{11}^2+Q_{12}^2)Q_{12}+2L\nabla^2Q_{12}+ 2c_0 P_1 P_2, \label{TDGL_lc2} \\
\frac{1}{\Gamma_P} \frac{\partial P_1}{\partial t}&=& [-\alpha(\rho)-\beta {\bf P}\cdot {\bf P}] P_1 - \frac{v_1}{2\rho_0}\nabla_x \rho + \frac{\lambda_1}{\Gamma_P} ({\bf P} \cdot \nabla)P_1 +\lambda_2 \nabla_x (|{\bf P}|^2) \nonumber \\
&& +\lambda_3  P_1(\nabla \cdot {\bf P}) + \kappa \nabla^2 P_1 +  2 c_0 (Q_{11}P_1+Q_{12}P_2), \label{eq_P1} \\
\frac{1}{\Gamma_P} \frac{\partial P_2}{\partial t}&=& [-\alpha(\rho)-\beta {\bf P}\cdot {\bf P}] P_2 - \frac{v_1}{2\rho_0}\nabla_y \rho + \frac{\lambda_1}{\Gamma_P} ({\bf P} \cdot \nabla)P_2 +\lambda_2 \nabla_y (|{\bf P}|^2) \nonumber \\
&& +\lambda_3  P_2(\nabla \cdot {\bf P}) + \kappa \nabla^2 P_2 +  2 c_0 (Q_{12}P_1-Q_{11}P_2), \label{eq_P2} \\
\frac{1}{\Gamma_\rho} \frac{\partial \rho}{\partial t}&=& -\frac{v_0}{\Gamma_\rho} \nabla \cdot({\bf P}\rho) + D_{\rho} \nabla^2 \rho. \label{eq_rho}
\end{eqnarray}

The $\pm$ signs in Eqs.~\eqref{TDGL_lc1}-\eqref{TDGL_lc2} refer to $T>T_c~(-)$ and $T<T_c~(+)$, where $T_c$ is the ordering temperature of the pure nematic. Notice that the free energy yields $\lambda_2=w/2$ and $\lambda_3=-w$ in these equations. However, both of these dynamical terms are permitted by symmetry considerations. Therefore, we treat $\lambda_2$ and $\lambda_3$ as unrelated phenomenological parameters.

\section{Fixed Point Solutions and Linear Stability Analysis}
\label{Appendix_2}

The dimensionless Eqs.~(10)-(14) govern the evolution of the LLC to its steady state. It is useful to study the fixed point (FP) solutions (${\bf Q^*},{\bf P^*}$), as these dictate the nature of the domains and steady states formed during the evolution. To determine the FP solutions for the coupled system, we set $\partial /\partial t = \nabla = 0$ in Eqs.~(10)-(14) with $\xi_1=\xi_2=1$:
\begin{eqnarray}
&& \pm Q^*_{11}-({Q^*_{11}}^2+{Q^*_{12}}^2)Q^*_{11}+c_0({P^*_1}^2-{P^*_2}^2)=0, \label{SS_1} \\
&& \pm Q^*_{12}-({Q^*_{11}}^2+{Q^*_{12}}^2)Q^*_{12}+2c_0 P^*_1 P^*_2=0, \label{SS_2}\\
&& (g_0-|{\bf P^*}|^2)P^*_1+c_0 (Q^*_{11}P^*_1+Q^*_{12}P^*_2)=0, \label{SS_3} \\
&& (g_0-|{\bf P}^*|^2)P^*_2+c_0 (Q^*_{12}P^*_1-Q^*_{11}P^*_2)=0, \label{SS_4}
\end{eqnarray}
where $g_0=\rho_0/\rho_c-1$. The conservation law dictates that the homogeneous FP solution of Eq.~(14) is $\rho=\rho_0$. A trivial solution for Eqs.~\eqref{SS_1}-\eqref{SS_4} is $Q_{11}^*=0$ , $Q_{12}^*=0$, $P_1^*=0$ , $P_2^*=0$, which corresponds to a disordered state for both components.

The non-trivial FPs are rotationally invariant and can be expressed as: 
\begin{equation}
\label{SS_case22}
Q_{11}^*= r_Q\cos{2\theta}, \quad Q_{12}^*= r_Q\sin{2\theta}; \quad P_1^*=r_P\cos{\theta}, \quad P_2^*=r_P\sin{\theta}.
\end{equation}
Here, $\theta$ is the arbitrary angle between ${\bf P}^* \parallel {\bf n}^*$ and the $x$-axis. We can choose $\theta=0$ without loss of generality. This choice of $\theta$ corresponds to $Q^*_{11}=r_Q,\ P^*_1=r_P$ and $Q^*_{12}=P^*_2=0$. The substitution of these values in Eqs.~\eqref{SS_1}-\eqref{SS_4} simplifies them to 
\begin{eqnarray}
&& -r_Q^3+(\pm 1+c_0^2)r_Q \pm c_0 |g_0| = 0, \label{SS_5} \\
&& r_P^2 = c_0 r_Q \pm |g_0|. \label{SS_6}
\end{eqnarray}
Here, the first $\pm$ sign in Eq.~\eqref{SS_5} signifies $T<T_c~(+)$ or $T>T_c~(-)$. The $\pm$ sign with $|g_0|$ is dictated by whether $\rho_0>\rho_c~(+$) or $\rho_0<\rho_c~(-$). We solved these equations for arbitrary values of $c_0$. The FPs thus obtained are given in Table~\ref{Tab_1} for all cases. 
\begin{center}
\begin{table}[H]
\centering
\begin{tabular}{|c|c|}
\hline
\textbf{Cases} & \makecell{\textbf{FP solutions} \\
($Q_{11}^*,Q_{12}^*,P_1^*,P_2^*$) = ($r_Q,0,r_P,0$)} \\ 
\hline
         
\makecell{{\it Case 1}\\($T>T_c,\ \rho_0=\rho_c^+$) } &  \makecell{$r_Q=-2^{1/3}(1+c_0^2)a_1^{-1/3}+a_1^{1/3}(2^{1/3}3)^{-1}$\\
$r_P^2=c_0r_Q+|g_0|$\\
$a_1=27|g_0|c_0+\sqrt{(27|g_0|c_0)^2+4(3-3c_0^2)^3}$}\\
\hline
              
\makecell{{\it Case 2}\\($T<T_c,\ \rho_0=\rho_c^-$) } &  \makecell{$r_Q=2^{1/3}(1+c_0^2)a_1^{-1/3}+a_1^{1/3}(2^{1/3}3)^{-1}$ \\
$r_P^2=c_0r_Q-|g_0|$ \\
$a_1=-27|g_0|c_0+\sqrt{(27|g_0|c_0)^2+4(3-3c_0^2)^3}$} \\
\hline

\makecell{{\it Case 3}\\($T<T_c,\ \rho_0=\rho_c^+$) } &  \makecell{$r_Q=2^{1/3}(1+c_0^2)a_1^{-1/3}+a_1^{1/3}(2^{1/3}3)^{-1}$ \\
$r_P^2=c_0r_Q+|g_0|$ \\
$a_1=27|g_0|c_0+\sqrt{(27|g_0|c_0)^2+4(3-3c_0^2)^3}$} \\
\hline
\end{tabular}
\caption{FP solutions for {\it Cases 1-3}.}
\label{Tab_1}
\end{table}
\end{center}

Next, we determine the stability of the FP solutions ($\rho_0, {\bf P^*, Q^*}$). The evolution of small fluctuations around these solutions ($\rho_0 + \Delta\rho, {\bf P^*} + \Delta {\bf P}, {\bf Q^*} + \Delta {\bf Q}$) can be obtained using Eqs.~(10)-(14). It is convenient to work with Fourier-transformed fluctuations
[$\Delta\rho({\bf k},t), \Delta{\bf P}({\bf k},t), \Delta{\bf Q}({\bf k},t)$]. The corresponding linearized equations can be written in vector notation:
\begin{equation}
\label{LS}
\frac{\partial \Phi ({\bf k},t)}{\partial t}= W({\bf k}) \cdot \Phi ({\bf k},t) ,
\end{equation}
where $\Phi({\bf k},t)=[\Delta \rho({\bf k},t), \Delta P_1({\bf k},t), \Delta P_2({\bf k},t), \Delta Q_{11}({\bf k},t), \Delta Q_{12}({\bf k},t)]$. The quantity $W({\bf k})$ is a $5\times 5$ matrix:

\begin{equation}
W=
\begin{pmatrix}
\makecell{iv_0^\prime(k_xP_1^* + k_yP_2^*) \\- D_{\rho}^\prime(k_x^2 + k_y^2)} & i k_x v_0^\prime\rho_0 & ik_yv_0^\prime\rho_0 & 0 & 0\\

%2nd
\dfrac{P_1^*}{\rho_c}+\dfrac{ik_xv_1^\prime}{2\rho_0} & \makecell {\dfrac{\rho_0}{\rho_c}-1 -3 {P_1^*}^2-{P_2^*}^2\\
- i k_x ({\lambda_1^\prime + 2 \lambda_2^\prime + \lambda_3^\prime)P_1^*}\\-i k_y \lambda_1^\prime P_2^*-\kappa^\prime(k_x^2 + k_y^2)\\ + c_0Q_{11}^*} & \makecell {- 2 P_1^* P_2^* -2i k_x \lambda_2^\prime P_2^*\\ - i k_y \lambda_3^\prime P_1^*   + c_0Q_{12}^*} & c_0 P_1^* & c_0 P_2^*\\

%3rd
\dfrac{P_2^*}{\rho_c}+\dfrac{ik_yv_1^\prime}{2\rho_0}&  \makecell {- 2P_1^* P_2^* -2i k_y\lambda_2^\prime P_1^*\\ - i k_x\lambda_3^\prime P_2^*+ c_0Q_{12}^*} & \makecell {\dfrac{\rho_0}{\rho_c}-1 -3 {P_2^*}^2-{P_1^*}^2\\
- i k_y ({\lambda_1^\prime + 2 \lambda_2^\prime + \lambda_3^\prime)P_2^*}\\-i k_x \lambda_1^\prime P_1^*-\kappa^\prime(k_x^2 + k_y^2)\\ - c_0Q_{11}^*} &-c_0 P_2^* &c_0 P_1^*\\

%4th
 0 & 2c_0 P_1^*  & -2c_0P_2^* & \makecell{\pm1-3{Q_{11}^*}^2 \\ -{Q_{12}^*}^2\\ -(k_x^2 + k_y^2)} & -2Q_{11}^*Q_{12}^* \\
 
 %5th 
 0 &  2c_0P_2^* & 2c_0P_1^* &-2Q_{11}^*Q_{12}^* & \makecell{\pm1 -3{Q_{12}^*}^2 \\-{Q_{11}^*}^2 \\-(k_x^2 + k_y^2)} 
\end{pmatrix}
\label{w}
\end{equation}

As usual, the eigenvalues $\{\bar{\lambda}_i\}$ and eigenvectors of $W({\bf k})$ determine the stability of a FP. If any ${\bar\lambda}_i > 0$, the fluctuations grow exponentially in time in the corresponding eigen-direction, i.e., the FP is unstable. To examine the stability of the disordered solution, we set $P_1^*=P_2^*=Q_{11}^*=Q_{12}^*=0$ in Eq.~\eqref{w}. It is clear that the coupling terms do not contribute at the linear level as they are quadratic in $P_i$ and $Q_{ij}$. Thus, the stability properties of the trivial disordered FP are the same as those of the LC and AM separately.

For non-trivial FPs, the analysis is more complicated and and analytically ugly even after setting $P_2^*=Q_{12}^*=0$. We determine the $\{{\bar\lambda}_i ({\bf k})\}$ numerically as a function of ${\bf k}$, and see whether any of the values lies above 0. For example, consider the phase diagram in Fig.~1(i). For large values of $\rho_0-\rho_c$, the system lies in the ordered state of {\it Case 1} in Table~\ref{Tab_1}. Thus, all eigenvalues are negative-definite for this state. We reduce the value of $\rho_0-\rho_c$ at constant $c_0$, and investigate where the first instability arises. This signals the onset of a non-trivial ordered state with spatial inhomogeneity, which is identified as a chimera. This is how the dashed lines in Fig.~1(i) and Fig.~3(a)-(b) are obtained.

In {\it Case 2}, we also have a non-trivial FP where $Q_{11}^*=1, Q_{12}^*=P_1^*=P_2^*=0$., i.e., the LC is ordered and AM is disordered. The dotted line in Fig.~3(a) denotes the boundary where this isotropic state becomes unstable, foreshadowing the onset of order in both fields.

\section{Movies Showing Steady States of LLCs}
\label{Appendix_3}

The movies below show the evolution of the active field (right frame) and nematic field (left frame) to different steady states from the initially disordered state. The steady states exist throughout the simulation time ($t=50000$), even though these are shown in the movies only up to time $t=2000$.
\begin{itemize}

\item \href{https://csciitd-my.sharepoint.com/:v:/g/personal/phz178374_iitd_ac_in/EYI3GlRZb9RDp7sqjTUcIzcBxpKHUeWNuRoJnC92WXTKfA?e=FMkPCF}{Movie 1}: Evolution of the LLC into a chimera for {\it Case 1}. The parameters are $T>T_c$ and $\rho_0=\rho_c^+=0.52$ with the coupling strength $c_0=0.5$.

\item \href{https://csciitd-my.sharepoint.com/:v:/g/personal/phz178374_iitd_ac_in/ETdIZQwFB1dLryRCmiB4a58BVWGvLloUl6vmxVLHevgzHg?e=XWrGAb}{Movie 2}: Evolution of the LLC to the chimera state for {\it Case 2}: $T<T_c$, $\rho_0=\rho_c^-=0.48$ with $c_0=0.1$. We point out here that the chimera in the nematic component manifests only after the annihilation of all defects (points of vanishing ${\mathcal S}$).

\item \href{https://csciitd-my.sharepoint.com/:v:/g/personal/phz178374_iitd_ac_in/EdjYg-PEhSNLuU-Q5nnZXycBrsBEgyGMDPNWVBL4OoV9tA?e=8TYPem}{Movie 3}: The 2-dimensional soliton for {\it Case 2}: $T<T_c$, $\rho_0=\rho_c^-=0.48$ with $c_0=0.1$. The nematic component exhibits the soliton only after the annihilation of all defects. 
    
\end{itemize}

\newpage

\bibliography{livelc.bib}
\bibliographystyle{apsrev4-1}

\end{document}